\documentclass[12pt]{article}

\makeatletter
\def\blfootnote{\xdef\@thefnmark{}\@footnotetext}
\makeatother
\usepackage{graphicx}

\usepackage{marvosym}

\newtheorem{proposition}{Proposition}

\newtheorem{corollary}{Corollary}
\newtheorem{example}{Example}

\usepackage[ruled]{algorithm2e}

\SetAlFnt{\small}
\SetAlCapFnt{\small}
\SetAlCapNameFnt{\small}
\SetAlCapHSkip{0pt}
\IncMargin{-\parindent}
\SetKwInOut{Input}{Input}
\SetKwInOut{Output}{Output} 

\usepackage{amsfonts,enumerate,subfigure,latexsym,amssymb,amsmath} 
\usepackage{fullpage}
\usepackage{tabularx, multirow, subfigure, rotating}
\usepackage{ifdraft}
\usepackage[table]{xcolor} 
\usepackage{color}
\usepackage{hyperref}
\hypersetup{
pdfborder={0 0 0 [0 0]},
pdftitle={Faster Geometric Algorithms via Dynamic Determinant Computation},
pdfauthor={V. Fisikopoulos, L. Pe{\~n}aranda},
pdfkeywords={determinant algorithms, Orientation predicate, volume
computation, rank-1 updates, experimental analysis}
}
\usepackage{xspace}

\newcommand\dyninv{{\it dyn\_inv}\xspace}
\newcommand\dynadj{{\it dyn\_adj}\xspace}
\newcommand\ie{\emph{i.e.}\xspace}
\newcommand\eg{\emph{e.g.}\xspace}
\newcommand\etal{\emph{et~al.}\xspace}

\def\RR{{\mathbb R}}  
\def\A{{\mathcal A}}

\newcommand{\nc}[1]{}
\newcommand{\nm}[1]{}

\newcounter{inlineequation}
\definecolor{highl}{rgb}{0.88,1,1}

\title{Faster Geometric Algorithms via Dynamic~Determinant~Computation}

\author{Vissarion~Fisikopoulos\textsuperscript{\Coffeecup}
\and Luis~Pe{\~n}aranda\textsuperscript{\Football}}

\newcommand{\mailto}[1]{\href{mailto:#1}{\texttt{#1}}}

\begin{document}

\maketitle
\blfootnote{\hspace{-3.8pt}\textsuperscript{\Coffeecup}Universit\'e Libre
de Bruxelles. \mailto{vfisikop@ulb.ac.be}}

\blfootnote{\hspace{-3.8pt}\textsuperscript{\Football}Universidade Federal
do Rio de Janeiro. \mailto{luis.penaranda@gmx.com}}

\begin{abstract}
The computation of determinants or their signs is the core procedure in
many important geometric algorithms, such as convex hull, volume and point
location. As the dimension of the computation space grows, a higher
percentage of the total computation time is consumed by these computations.
In this paper we study the sequences of determinants that appear in
geometric algorithms. The computation of a single determinant is
accelerated by using the information from the previous computations in that
sequence.

We propose two dynamic determinant algorithms with qua\-dratic arithmetic
complexity when employed in convex hull and volume computations, and with
linear arithmetic complexity when used in point location problems. We
implement the proposed algorithms and perform an extensive experimental
analysis. On one hand, our analysis serves as a performance study of
state-of-the-art determinant algorithms and implementations. On the other
hand, we demonstrate the supremacy of our methods over state-of-the-art
implementations of determinant and geometric algorithms. Our experimental
results include a $20$ and $78$ times speed-up in volume and point location
computations in dimension $6$ and $11$ respectively.

\vspace{4mm}
\end{abstract}

\section{Introduction}\label{sect:intro}

Computing the sign of a determinant, or in other words evaluating a determinant {\it predicate}, is in the core of many important geometric algorithms. For example, convex hull
algorithms use \emph{orientation} predicates, and Delaunay
tri\-an\-gu\-la\-tion algorithms involve \emph{in-sphere} predicates. 
Furthermore, the computation of the value of a determinant, or in other
words a determinant {\it construction}, is also important in some geometric
algorithms. For example, the exact volume computation of a convex polytope
using either triangulation or sign decomposition method relies on the
computation of the volume of simplices, which reduces to computing the value of a determinant.

In other words predicates encapsulate decisions in contrast to constructions that involve computation of new numerical values. 
In general dimension $d$, the orientation predicate of $d+1$ points is the sign of the determinant of a matrix containing the homogeneous coordinates of the points as columns.
On the other hand, the volume of a simplex is the value of the  determinant of a matrix containing the homogeneous coordinates of the $d+1$ vertices of the simplex.
In practice, as the dimension grows, a higher percentage of the computation time is consumed by these core procedures.

In this paper, we study effective algorithms and implementations for the computation of the determinant predicates and  constructions that appear in geometric computations. 
The model we follow is the exact computation paradigm presented in~\cite{Yap_exact} and advocated by the Computation Geometry Algorithms Library ({\tt CGAL})~\cite{CGAL}, a state-of-the-art library for geometric computations. Note that in geometric algorithms the naive use of floating point arithmetic may lead to incorrect results~\cite{KMPSY08}.
There are two main scenarios regarding exactness. The first provides exact
predicates but not necessarily exact constructions while the second
provides both exact predicates and exact constructions. In this paper we
study the second scenario. 
We give a particular emphasis on exact division and division-free
algorithms. Avoiding divisions is crucial when working on a ring that is
not a field, \eg, integers or polynomials.

The main idea of our approach is to study the {\it sequence} of
computations of determinants or signs of determinants that appear in geometric algorithms. A single 
computation can be accelerated by using the information from the previous computations in this sequence.
The essential case is the sequence of computations of the  orientation predicates that appear in convex hull algorithms.
The convex hull problem is probably the most fundamental problem in discrete and computational geometry. In fact, the problems of regular,   
Delaunay triangulations and Voronoi diagrams reduce to it by computing a 
convex hull in one dimension higher~\cite{Bois_book}. Additionally, in the course of an incremental convex hull algorithm like Beneath-and-Beyond~\cite{Seidel81} we compute the volume of the polytope as a by-product of the computation. See~\cite{Bueler98exactvolume} for a survey on volume computation and relevant implementations.

Since we will study {\it in practice} the performance of geometric and
algebraic algorithms, it is important to classify the test cases.
Especially, one of the parameters we will use is the dimension. We will
refer throughout the paper to dimensions $d<5$ as \emph{low}, to dimensions
$5\le d\le 25$ as \emph{medium} and to dimensions $d>25$ as \emph{high},
unless otherwise stated.
In our experiments, we focus on medium dimensions for determinant
computations and ``small to medium" for geometric algorithms, \ie, 6 to 11 depending on the application. 

\subsection{Previous work}
There is a variety of algorithms and implementations for computing the determinant of a $d\times d$ matrix. 
Let us denote by $O(d^{\omega})$ the complexity of matrix multiplication. 
First, we consider the case where the matrix has values from a field. 
For $\omega>2$, an algorithm for matrix multiplication imply an algorithm for determinant computation with the same $\omega$~\cite{BH74}. 
The best current  $\omega$ is $2.3728639$~\cite{LeGall14}.

An important class of determinant computation algorithms are the algorithms
which use {\it exact divisions}, \ie, divisions known to have remainder
zero.
An application of them is the computation of the determinant of a matrix
with integer entries using only integer arithmetic. A typical example of
this is Bareiss algorithm~\cite{Bareiss68}. 

{\it Division-free} algorithms form another category. They use no
divisions at all, \eg, when matrix coefficients are elements of an abstract
commutative ring.
The best current $\omega$ in this category is \(2.697263\)~\cite{KaVi05}.  Here,
it is worth mentioning a family of determinant algorithms that use combinatorial
approaches.
They were introduced by Mahajan and Vinay~\cite{MV97}, and are based on
\emph{clow} (closed ordered walk) sequences.
Several similar methods with complexity \(O(d^4)\) are surveyed by
Rote~\cite{Rote01}.
Based on the idea of clow sequences Bird introduced a simpler algorithm that
uses matrix operations~\cite{Bird11}.
Its complexity is \(O(dM(d))\), where \(M(d)\) is the complexity of matrix
multiplication.
Urba{\'n}ska conceived a method that uses fast matrix multiplication~\cite{CW87}
to obtain a complexity \(O(d^{3.03})\)~\cite{Urb10}.
However, in practice when $d$ is small, Bird's algorithm behaves better than other division-free algorithms, as it will be discussed 
in Section~\ref{subsec:dce}.

Determinants of matrices over a ring arise in combinatorial problems~\cite{krattenthaler05}, in algorithms for lattice polyhedra~\cite{Barvinok99analgorithmic} and secondary polytopes~\cite{RambTOPCOM} or in computational algebraic geometry problems~\cite{Cox-Little-OShea:UAG2005}. A special case of the latter is the computation of resultant polytopes that have applications in polynomial system solving~\cite{BaPoRo} and geometric modeling~\cite{EKKL12}.

Good asymptotic complexity does not imply good behavior in practice for low and medium dimensions.
For instance, {\tt LinBox}~\cite{DGGGHKSTV}, which implements algorithms
with state-of-the-art asymptotic complexity, introduces a significant
overhead in low and medium dimensions, and seems most suitable in high
dimensions (see Section~\ref{subsec:dce} for more details).

{\tt Eigen}~\cite{eigenweb}
implements LU~decomposition, of complexity \(O(d^3)\), and seems to be
suitable for low and medium dimensions. {\tt Eigen} was designed with
floating-point computations in mind, where it uses hardware floating-point
vectors to attain great speed.

In addition, there exists a variety of algorithms for determinant sign
computation~\cite{C92,BEPP99,AbBrMu99,BY00,BroBurPio01}. Kaltofen and
Villard~\cite{KV04} present a complete survey on the matter. One tool
commonly used for sign computations is {\it filtering}: arithmetic
operations are done using fixed-precision floating-point interval
arithmetic, switching to exact arithmetic only when the sign is unknown.
Filtered computations are widely used because they provide a simple
approach to avoid performing exact operations in many cases. While filtered computation performs well in low dimensions, there is no experimental study on the efficiency of current methods in medium dimensions (see Section~\ref{subsec:fil}).

The problem of computing sequences of determinants has also been studied.
TOPCOM~\cite{RambTOPCOM} is the reference software for
enumerating all regular triangulations of a set of points in general  dimension. It efficiently pre-computes all
orientation determinants that will be needed in the computation and stores their signs.
Emiris~\etal~\cite{EFKP12journal} study a similar problem in the context of
computational algebraic geometry. In particular, the computation of several
regular triangulations for different lifting functions. The computation of
orientation predicates is accelerated by maintaining a hash table with the
computed minors of the determinants. These minors appear many times in the
computation. 
However, this method does not provide considerable acceleration when
applied to the case of a single convex hull computation.

Our approach utilizes the Sherman--Morrison formulas~\cite{SM50,Bart51}. 
They relate the inverse of a matrix after a
small-rank perturbation to the inverse of the original matrix.
Other applications of these formulas include solving the dynamic transitive
closure problem in graphs~\cite{Sankowski04} and studying the effect of new
links on Google Page Rank~\cite{Avrachenkov06theeffect}.

\subsection{Contribution} 
We design algorithms that perform dynamic determinant updates and achieve qua\-dratic complexity for the determinants involved in  incremental convex hull or volume computation algorithms and linear
complexity for determinants involved in point location algorithms. Interestingly, we propose a variant of these algorithms that can perform computations over the integers. Our main technical tool is Sherman--Morrison formulas. As far as we know this is the first application of these formulas to geometric algorithms.  

We implement the proposed algorithms along with division-free
determinant algorithms from the literature. We perform an extensive experimental analysis of the current state-of-the-art packages for exact determinant computations along with our implementations. Without taking the dynamic determinant algorithms into account, our experiments present a result of independent interest: they serve as a {\it survey} of state-of-the-art determinant algorithms and implementations.
In the division-free case, with matrices containing very large integer
values, we show that the simple and not-widely used algorithm due to
Bird~\cite{Bird11} outperforms state-of-the-art implementations in
dimensions $6<d<10$, while providing a very competitive
performance for higher dimensions.
Dynamic algorithms start outperform all
the other tested determinant implementations in dimension $6$ when the
input has small bit-size. For larger bit-size, the dynamic algorithms become competitive in larger dimensions ($d>23$ in our tests).

We adapt our implementations to work with geometric algorithms, thus providing exact predicates and constructions.
A natural geometric context to test our method is exact volume computation,
where it yields very competitive implementations. For instance, we obtain
an up to $20$ times speed-up comparing with state-of-the-art packages in
dimension $6$.
We also provide experimental results showing that our method improves the
running time of convex hull and point location implementations with respect
to other exact implementations. 
Another interesting feature of our method is that it takes advantage of
multiple precision integer, as opposed to rational, arithmetic when the input
coordinates are integral (\eg lattice polytopes).

\paragraph{ Overview of the paper}
The paper is organized as follows.
Section~\ref{sect:dyndet} introduces the dynamic determinant algorithms and the following section presents their application to geometric algorithms. 
Section~\ref{sect:experiments} discusses the implementation, experiments, and comparison with other software.
We end up with conclusions and future work.

A preliminary version of the results of this paper appeared~\cite{FP_ESA12}. In this
final version, we include new results on volume
computations and experiments on real practical scenarios.
We also present more experimental results on determinant and convex hull computations and discuss issues as filtering and memory consumption.  
Overall, we provide an improved and more detailed presentation of our method.

\section{Dynamic Determinant Computations}\label{sect:dyndet}

In the \textit{dynamic determinant problem}, a $d\times d$ matrix $A$ is
given. Allowing some preprocessing, we should be able to handle
updates of elements of $A$ and return the current value of the determinant.
We consider here only non-singular updates, that is, updates that do not make $A$ singular. This assumption is sufficient for our method as we explain at the end of Section~\ref{sec:ch}.

The Sherman--Morrison formula~\cite{SM50,Bart51} states that
\begin{equation}\label{eq:init}
\left(A+wv^T\right)^{-1}=A^{-1}-\frac{(A^{-1}w)(v^TA^{-1})}{1+v^TA^{-1}w},
\end{equation}
where $A$ a $d\times d$ matrix and $v,w$ vectors of dimension $d$. 
Let \(A'\) be the matrix resulting from replacing the $i$-th column of $A$
by a vector $u$.
Also let \((A)_i\) denote the $i$-th column of $A$, and $e_i$ the vector with $1$ in its $i$-th place and $0$ everywhere else.
An $i$-th column update of $A$ is performed by substituting $v=e_i$ and
$w=u-(A)_i$ in Equation~\ref{eq:init}. 
Then, we can write $A'^{-1}$ as follows: 
\begin{align}
 A'^{-1}=\left(A+(u-(A)_i)e_i^T\right)^{-1}
        =A^{-1}-\frac{\big(A^{-1}(u-(A)_i)\big)\ (e_i^TA^{-1})}{
                1+e_i^TA^ { -1 }(u-(A)_i)}, \label{eq:inv}
\end{align}
where $e_i^T$ is simply selecting row $i$.
If $A^{-1}$ is computed, we compute $A'^{-1}$ using Equation~\ref{eq:inv}. 
The computation is performed as follows: 
\begin{align}
h_1&=A^{-1}(u-(A)_i)\label{eq1}\\
h_2&=h_1/(1+(h_1)^i)\label{eq2}\\
H_3&=h_2\,(A^{-1})^i\label{eq3}\\
A'^{-1}&=A^{-1}-H_3\label{eq4}
\end{align}
where $(A)^i,(h_1)^i$ denote the $i$-th row of $A$ and the $i$-th element $h_1$ respectively. The intermediate results are the $d$-dimensional vectors $h_1,h_2$ and the $d\times d$ matrix $H_3$. Hence, the equations~\ref{eq1},~\ref{eq2},~\ref{eq3},~\ref{eq4} are computed in $d^2+d,\,d+O(1),\,d^2,\,d^2$ arithmetic operations respectively and thus $3d^2+2d+O(1)$ in total. 

The matrix determinant lemma~\cite{Harv97} states that 
\begin{equation}\label{eq:mdl}
\det\left(A+wv^T\right) = \left(1 + v^TA^{-1}w\right)\,\det\left(A\right)
\end{equation}
which yields the following equation 
\begin{align}
\det\left(A'\right)=\det\left(A+(u-(A)_i)e_i^T\right)
        =\left(1+e_i^TA^{-1}(u-(A)_i)\right) \det\left(A\right).
        \label{eq:detA}
\end{align}
Using Equation~\ref{eq:detA} we compute $\det\left(A'\right)$ in
\(2d+O(1)\) arithmetic operations, if $\det\left(A\right)$ is known.
Equations~\ref{eq:inv} and \ref{eq:detA} lead to the following result.

\begin{proposition}{\rm\cite{SM50}}\label{dynamic_det}
The dynamic determinant problem can be solved using $O(d^\omega)$ arithmetic
operations for preprocessing and $O(d^2)$ for non-singular one column
updates. The preprocessing consists in the computation of $A^{-1}$ and
$\det\left(A\right)$.
\end{proposition}

Then we show how this computation can be performed over a ring.
To this end, we use the adjoint of $A$, denoted by $A^{\text{adj}}$, rather than
the inverse. It holds that $A^{\text{adj}}=\det(A)A^{-1}$, thus we obtain
the following two equations.
\begin{align}
A'^{\text{adj}}&=\frac{1}{\det(A)}
        \Big(A^{\text{adj}}\det(A')-\left(A^{\text{adj}}(u-(A)_i)\right)
        \ \left(e_i^TA^{\text{adj}}\right)\Big)\label{eq:adj_ring}\\
\det(A')&=\det(A)+e_i^TA^{\text{adj}}(u-(A)_i)\label{eq:detA_ring} 
\end{align}
The only division in Equation~\ref{eq:adj_ring} is known to be exact,
\ie, its remainder is zero.
If the computation follows the order of operations as determined by the parenthesis in Equations~\ref{eq:adj_ring},~\ref{eq:detA_ring} 
then the computation can be performed in $5d^2+d+O(1)$ arithmetic operations for Equation~\ref{eq:adj_ring} and in \(2d+O(1)\) for Equation~\ref{eq:detA_ring}.
In the sequel, we will call \dyninv the dynamic determinant algorithm that uses Equations~\ref{eq:inv} and \ref{eq:detA}, and \dynadj the
one that uses Equations~\ref{eq:adj_ring} and \ref{eq:detA_ring}.

\section{Geometric Algorithms}\label{sect:convex_hulls}

We introduce in this section our methods for optimizing the computation of sequences of determinants that appear in geometric algorithms.
First, we utilize dynamic determinants, as described in the previous section, in incremental convex hull 
algorithms; they form one of the basic classes of convex hull algorithms.
Then, we show how this solution can be extended to other geometric algorithms such as point locations in triangulations and volume computations. 

\subsection{Preliminaries} 
Let us start with some basic definitions from discrete geometry. 
Let $\A\subset\RR^d$ be a set of $n$ points.
We define the \textit{convex hull} of a pointset $\A$, denoted by
$\text{conv}(\A)$, as the smallest convex set containing $\A$.
A hyperplane \textit{supports} $\text{conv}(\A)$ if $\text{conv}(\A)$ is
entirely contained in one of the two closed half-spaces determined by the
hyperplane and has at least one point on the hyperplane.
A \textit{face} of $\text{conv}(\A)$ is the intersection of
$\text{conv}(\A)$ with a supporting hyperplane that does not contain
$\text{conv}(\A)$.
Faces of dimension $0$ and $d-1$ are called {\it vertices} and {\it facets} respectively.
We call a face $f$ of $\text{conv}(\A)$ \textit{visible} from $a\in\RR^d$
if there is a supporting hyperplane that contains $f$ such that $\text{conv}(\A)$ is
contained in one of the two closed half-spaces determined by the hyperplane
and $a$ in the other.
A $k$-simplex of $\A$ is the convex hull of an affinely independent subset \(S\) of
$\A$, where $\text{dim}(\text{conv}(S))=k$.
A \textit{triangulation} of $\A$ is
a collection of simplices of $\A$, called the \emph{cells} of the triangulation,
such that the union of the simplices equals $\text{conv}(\A)$ and 
every pair of simplices intersect at a common face or have an empty intersection.
We define the {\it orientation matrix} $A_C$ of a set $C$ of points $\{a_1\dots a_{d+1}\}\subset \RR^d$ to be the $(d+1)\times (d+1)$ matrix such that for every $a_i$, the column $i$ of $A_C$ contains $\vec{a}_i$'s coordinates as entries, where $\vec{a_i}$ is the homogeneous vector $(a_i,1)$.

\subsection{Incremental convex hull}\label{sec:ch}

\begin{figure*}[t]
\centering
\includegraphics[width=0.531\textwidth]{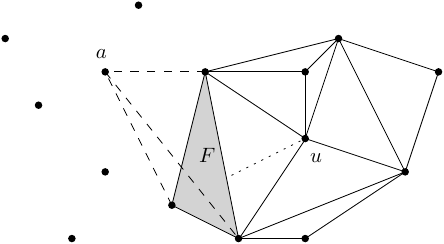}
\caption{The course of an incremental convex hull algorithm in 3 dimensions.\label{fig:ch_examples}}
\end{figure*}

\begin{algorithm}[t]
  \BlankLine
  \Input{pointset $\A\subset\RR^d$}
  \Output{convex hull of $\A$}
  \BlankLine
  sort $\A$ by increasing lexicographic order of coordinates, \ie,
    $\A=\{a_1,\dots,a_n\}$\;
  $T\leftarrow \{a_1,\dots,a_{d+1}\}$\;
  $Q\leftarrow$ facets of $\text{conv}(a_1,\dots,a_{d+1})$\;
  \BlankLine
  \ForEach{$a\in\{a_{d+2},\dots,a_n\}$ }{
	$Q' \leftarrow Q$\;
        \BlankLine
	\ForEach{$F\in Q$}{
	  $C \leftarrow$ the unique $d$-face s.t. $C\in T$ and $F\in C$\; 
	  $u \leftarrow$ the unique vertex s.t. $u\in C$ and $u\notin F$\;
	  $C'\leftarrow F \cup \{a\}$\;
          \tcp{$\det(A_{C})$ and $A^{\text{\textrm{adj}}}$ were computed in
          a previous step}
          $\det(A_{C'})\leftarrow$ ($\det(A_{C})$ after updating $u$ with $a$ using 
              Equations~\ref{eq:adj_ring}, \ref{eq:detA_ring})\; 
	  \BlankLine
          \If{$\det(A_{C'})\det(A_{C})<0$}{ 
	    $T\leftarrow T\cup \{\text{$d$-face of conv}(C')\}$\;
	    $Q'\leftarrow Q'\, \ominus\, \{(d-1)\text{-faces of }C'\}$;\, 
	    \tcp{symmetric difference}
	  }
       }
       $Q \leftarrow Q'$\;
  }
  \BlankLine
  \Return $Q$\;
  \caption{\label{AlgBB} Incremental Convex Hull\, ($\A$)}
\end{algorithm}

For simplicity, we assume general position of $\A$ and present our method for the
Beneath-and-Beyond (BB) algorithm~\cite{Seidel81}.
However, our method can be extended to handle degenerate
inputs as in~\cite[\S 8.4]{Edelsb87}, and can be applied to more efficient 
incremental convex hull algorithms (\eg,~\cite{CS89}) by utilizing the dynamic determinant computations to answer the predicates appearing in point location 
(Corollary~\ref{thm:point_location}).
A clarification of this claim is our implementation in 
Section~\ref{sect:experiments} which first handles degenerate inputs in practice and second is faster compared to other software.    
In what follows, we use the dynamic determinant algorithm \dynadj,
which can be replaced by \dyninv yielding a variant of the presented
convex hull algorithm.
This choice is supported by our experiments where we show that \dynadj is faster than \dyninv in all the tested dimensions. 

The BB algorithm is initialized by computing a $d$-simplex of $\A$. At
every subsequent step, a new point from $\A$ is inserted, while keeping a
triangulated convex hull of the inserted points. Let $t$ be the number  of
cells of this triangulation. Assume that, at some step, a new point
$a\in\A$ is inserted and $T$ is the triangulation of the convex hull of the
points of $\A$ inserted up to now. To determine if a facet $F$ is visible
from $a$, an orientation predicate involving $a$ and the vertices of $F$
has to be computed (Figure~\ref{fig:ch_examples}). That is, we have to
compute the sign of the determinant of the matrix $A_{C}$, where $C$ is the
set of vertices of $F$ union with $a$. If we know the adjoint and the
determinant of the orientation matrix of a cell of $T$ that contains $F$,
this can be done by applying  Equation~\ref{eq:detA_ring}. If $F$ is on the
boundary, this cell is unique (\eg, $(F,u)$ in Figure~\ref{fig:ch_examples}) otherwise we arbitrarily select one of the two cells that contain $F$. 

Algorithm~\ref{AlgBB}, as initialization, computes from
scratch the adjoint matrix and the determinant of the orientation matrix $A_C$, where $C$ contains the vertices of the initial $d$-simplex.
At every incremental step, it first computes the orientation predicates using the
adjoint matrices and determinants computed in previous steps using  Equation~\ref{eq:detA_ring}. Second, it computes the adjoint and determinant of the orientation matrices of the new cells using Equation~\ref{eq:adj_ring}.
By Proposition~\ref{dynamic_det}, this method leads to the following result. 

\begin{proposition}\label{thm:ch}
Given a $d$-dimensional pointset the first orientation predicate of incremental convex hull algorithms is computed in $O(d^\omega)$ time, and all the others  
in $O(d^2)$ time in total $O(d^2t)$ space, where $t$ is the number of cells of the constructed triangulation.
\end{proposition}

Essentially, this result improves the computational complexity of the
predicates involved in incremental convex hull algorithms from
$O(d^{\omega})$ to $O(d^2)$ by using more space and dynamic determinant updates.
Recall that $O(d^{\omega})$ is the current best complexity (Section~\ref{sect:intro}).
To analyze the complexity of Algorithm~\ref{AlgBB}, we bound the number of
facets of $Q$ in every step of the outer loop of Algorithm~\ref{AlgBB} with
the number of $(d-1)$-faces of the constructed triangulation of
$\text{conv}(\A)$, which is bounded by $(d+1)t$. Thus, using
Lemma~\ref{thm:ch}, we have the following complexity bound for
Algorithm~\ref{AlgBB}, where we assume that $n\gg d$ to hide the preprocessing complexity $O(d^{\omega})$.

The method of dynamic determinants increases the {\it space complexity} of BB from $O( n d)$ numbers and $O(td)$ references to $O( t d^2)$ numbers and $O(td)$ references. 
The numbers stored by the two methods are different. The original method stores only point coefficients while ours stores additionally determinants and the inverse and adjoint matrices. 
The bit-sizes of those numbers are different. The $O( n d)$ point
coefficients are part of the input. Let $\tau$ be a  bound on their
bit-sizes. From Hadamard's inequality~\cite{Garling07} the value of the
determinant of a matrix $A$ is bounded by
\[|\det(A)|\leq 2^{\tau d} d^{d/2}\text{.}\]
It follows that the bit-size of the computed determinants is $O(d(\tau+\log
d)$, which becomes $O(d\tau)$ under the standard assumption $\tau \gg d$.
Since the absolute values of the elements of the adjoint and inverse of $A$
are bounded by the determinant of submatrices of $A$, the above bound also
holds for the bit-size of the elements of the adjoint and inverse matrices.

\begin{corollary}\label{cor:1}
Given $n$ $d$-dimensional points whose coefficients bit-size is bounded by
$\tau$, the complexity of BB algorithm is $O(n\log n+d^3nt)$, where $n\gg
d$, $\tau \gg d$ and $t$ is the number of cells of the constructed
triangulation. The consumed space is $O( t d^2)$ numbers of bit-size at
most $O(d\tau)$ and $O(td)$ references.
\end{corollary}

Note that the complexity of BB, without using the method of dynamic
determinants, is bounded by $O(n\log n+d^{{\omega}+1}nt)$. Recall that $t$
is bounded by $O(n^{\lfloor d/2\rfloor})$~\cite[\S 8.4]{Ziegler}, which
shows that Algorithm~\ref{AlgBB}, and convex hull algorithms in general, do not
have polynomial complexity in $n$ and $d$.
The schematic description of Algorithm~\ref{AlgBB} and its coarse analysis is
good
enough for our purpose: to elucidate the application of dynamic
determinants  to incremental convex hull computation and to quantify the
improvements using this method. See Section~\ref{sect:experiments} for a
practical approach to incremental convex hull algorithms using dynamic
determinant computations.

In Section~\ref{sect:dyndet} we have addressed only non-singular updates.
Here we show that this will not limit our method to handle degenerate
cases. In a degenerate case, the determinant of an orientation matrix will
be zero if the points in the orientation test span a space of dimension
less than $d$. However, in this case, we do not have to update the adjoint
or the determinant of the orientation matrix (which would be equivalent to
a singular update operation), since no new cell is going to be created.  

\subsection{Point location and volume computation}
The above results can be used to improve the efficiency of geometric
algorithms that use convex hull computations.
One way of computing {\it Delaunay triangulations} in $\RR^{d}$ and their
dual {\it Voronoi diagrams} is to compute the convex hull of the points
lifted on the paraboloid in $\RR^{d+1}$. 
For generic liftings, the above construction leads to regular triangulations. 

Another important geometric problem where our method could be applied is {\it exact  volume computation}, since one of the two major classes of volume computation algorithms is based on 
triangulation methods~\cite{Bueler98exactvolume}.
To elucidate this, observe that in Algorithm~\ref{AlgBB} we can compute the
volume of the polytope by summing up the volumes of all full dimensional
simplices in the resulting triangulation. Indeed, the volume of a simplex
is the absolute value of the determinant of its orientation matrix. The
difference of an incremental convex hull and a volume computation algorithm
using a triangulation method is that the former needs to evaluate
determinant predicates (\ie, know only the sign of determinants), while the
latter needs determinant constructions (\ie, compute the value of
determinants).

As mentioned above, more efficient incremental convex hull algorithms
(\eg,~the work of Clarkson and Shor~\cite{CS89}) do not sort the input
points, they use instead {\it point location} methods to find the position
of the point that is going to be inserted into the convex hull.
It is straightforward to apply our scheme in orientation predicates
appearing in {\it point location} algorithms, that perform orientation
tests with respect to the facets of the triangulation.
The orientation predicates queried by a point location algorithm can be
computed using Equation~\ref{eq:detA_ring}, if the adjoint and determinant
of the orientation matrices of the cells of the triangulation have been
precomputed.
That yields the following result.

\begin{corollary}\label{thm:point_location}
Given a triangulation of a $d$-dimensional pointset computed by an
incremental convex hull algorithm like Algorithm~\ref{AlgBB}, the
orientation predicates involved in point location algorithms that perform
orientation tests with respect to the facets of the triangulation can be
computed in $O(d)$ arithmetic operations, using $O(d^2t)$ numbers of
maximum bit-size $O(d\tau)$ as space, where $t$ is the number of cells of
the triangulation and $\tau$ bounds the bit-sizes of the numbers, as in
Corollary~\ref{cor:1}.
\end{corollary}

\section{Implementation and Experimental Analysis}\label{sect:experiments}

\subsection{Software design}
We implemented in C++ the methodology described above, which we call {\it hashed dynamic determinants}.
The scheme consists of efficient implementations of algorithms \dyninv and \dynadj (Section~\ref{sect:dyndet}) and a hash table, which stores intermediate results such as matrices and determinants.
Note that since our implementation computes values of determinants  and not only their sign it cannot take advantage of filtering techniques (Section~\ref{subsec:fil}). 

The design of our implementation is {\it modular}. It can be used by an
algebraic software, providing dynamic determinant algorithm implementations.
Moreover it can be used by a geometric software providing exact geometric
predicates and constructions (\eg, orientation and volume). Here we focus
on geometric software that implements incremental convex hull algorithms,
which essentially compute a triangulation. Our implementation is
independent of the data-structures used by the geometric software. The use
of the hash table as an additional data-structure is a way to provide the
user with an interface to the new determinant computation without modifying
its own data structure.
In practice, hash tables have constant insertion and retrieval times, and
thus our approach does not introduce a significant overhead in computing
time while remaining modular.

The hashing scheme works as follows.  
Assume that the input points are indexed as $\{a_1,\dots,a_n\}$.
We use as {\it hash keys} the tuples of indices of the $(d-1)$-faces of the
triangulation. Each $(d-1)$-face is mapped to one of the two cells (\ie, $d$-faces) of the triangulation that it belongs to. The selection between the two cells is arbitrary and does not affect the efficiency of the method. For every cell we also store the adjoint and the determinant of the matrix that corresponds to its vertices' coordinates.  
In the course of geometric algorithms a given point $b$ should be tested for orientation with respect to a hyperplane defined by points that are locally indexed as  $a_1,\dots,a_d$. Querying the hash table for the tuple $(a_1,\dots,a_d)$ we obtain the adjoint and the determinant of the matrix with entries the coordinates of $a_1,\dots,a_d$ and one more point $c$. Thus, the requested orientation determinant is computed by updating $c$ with $b$ applying  Equations~\ref{eq:adj_ring} and \ref{eq:detA_ring}.  
The following 2-dimensional example illustrates our approach.

\begin{example}
Let $A=\{a_1=(0,1),\,a_2=(1,2),\,a_3=(2,1),\,a_4=(1,0),\,a_5=(2,2)\}$ where
every point $a_i$ has an index $i$ from $1$ to $5$.
 Assume we are in some step of an incremental convex hull or point location
algorithm and let $T=\{\{1,2,4\},\,\{2,3,4\}\}$ be the $2$-dimensional
triangulation
of conv($A$) computed so far. The cells of $T$ are indexed 
using the indices of the points in $\A$. 
For each cell, the hash table will store as keys the
set of indices of the $2$-faces of the cell, \eg,
for the cells $\{\{1,2,4\}$ the keys are $\{\{1,2\},\{2,4\},\{1,4\}\}$
mapping to the adjoint and the determinant of  the matrix constructed by the
points $a_1,a_2,a_4$. Similarly, $\{\{2,3\},\{3,4\},\{2,4\}\}$ are mapped to the
adjoint matrix and determinant of $a_2,a_3,a_4$. 
To insert $a_5$ in $T$ one should compute the orientation determinant of $a_2,a_3,a_5$ to determine whether the facet $\{2,3\}$ is visible from $a_5$ and hence should be connected to construct a new cell $\{2,3,5\}$. 
Similar computations are performed for the other facets. 
By querying
the hash table for $\{2,3\}$ the adjoint and the determinant of the matrix of
$a_2,a_3,a_4$ are returned. Then, we perform an update of the column
corresponding to point $a_4$, replacing it by $a_5$ and apply
Equations~\ref{eq:adj_ring} and \ref{eq:detA_ring} to compute the adjoint and the determinant of the new cell. Finally, the two new keys $\{2,5\},\{3,5\}$ are added to the hash table and are mapped to the new cell $\{2,3,5\}$. 
\end{example}

The hash table has been implemented using the Boost libraries~\cite{boost}.
To reduce memory consumption and speed-up look-up time, we sort the lists of
indices that form the hash keys.
We use the \emph{GNU Multiple Precision arithmetic library} (GMP), the
current standard for multiple-precision arithmetic, which provides integer
and rational types {\tt mpz\_t} and {\tt mpq\_t}, respectively.

The geometric software we interface with our implementation is the {\tt
CGAL} package {\tt Tri\-an\-gu\-la\-tion}~\cite{Hornus15,BoiDevHor09},
which implements an incremental convex hull algorithm. The
difference between this implementation and Algorithm~\ref{AlgBB} of Section~\ref{sect:convex_hulls} is that {\tt Triangulation} does not sort
the points along one coordinate but along a $d$-dimensional Hilbert curve
and performs a fast point location at every insertion. Thus, we can take
advantage of our scheme in two places: ({\it a}) in the orientation
predicates appearing in the {\it point location} procedure, and ({\it b})
in the ones that appear in the {\it construction of the convex hull}. 

We call {\tt hdch} the modification of {\tt Triangulation} with hashed dynamic determinants. 
On the technical part, we provide a modification of the {\tt CGAL} Kernel were the call to the determinant is replaced by a functor which implements
the dynamic determinant formulas and has access to the hash table. The hash
table is completely hidden from the interface.
We use {\tt Eigen} for initial determinant and adjoint or inverse matrix computation and Laplace determinant algorithm for dimensions lower than $6$. 

\subsection{Experimental setup}
All experiments ran on an Intel Core i5-2400 $3.1$GHz, with $6$MB L2 cache and
$8$GB RAM, running 64-bit Debian GNU/Linux.
We divide our tests in four scenarios, according to the number type
involved in computations:
\begin{description}
\item[a.]  rationals where the bit-size of both
numerator and denominator is $10000$,
\item[b.] rationals converted from {\tt doubles}, that is, numbers of the
form $m\times 2^p$, where $m$ and $p$ are integers of bit-size $53$ and $11$
respectively,
\item[c.] integers with bit-size $10000$, and
\item[d.] integers with bit-size $32$.
\end{description}
However, it is rare to find in practice input coefficients of scenarios (a)
and (c). Inputs are usually given as $32$ or $64$-bit numbers. These
inputs
correspond to the coefficients of scenario~(b). Scenario~(d) is also very
important, since points with integer coefficients are encountered in many
combinatorial applications (Section~\ref{sect:intro}).

\subsection{Determinant computation experiments}
\label{subsec:dce}

\begin{figure*}[t]
\centering
\includegraphics[width=\textwidth]{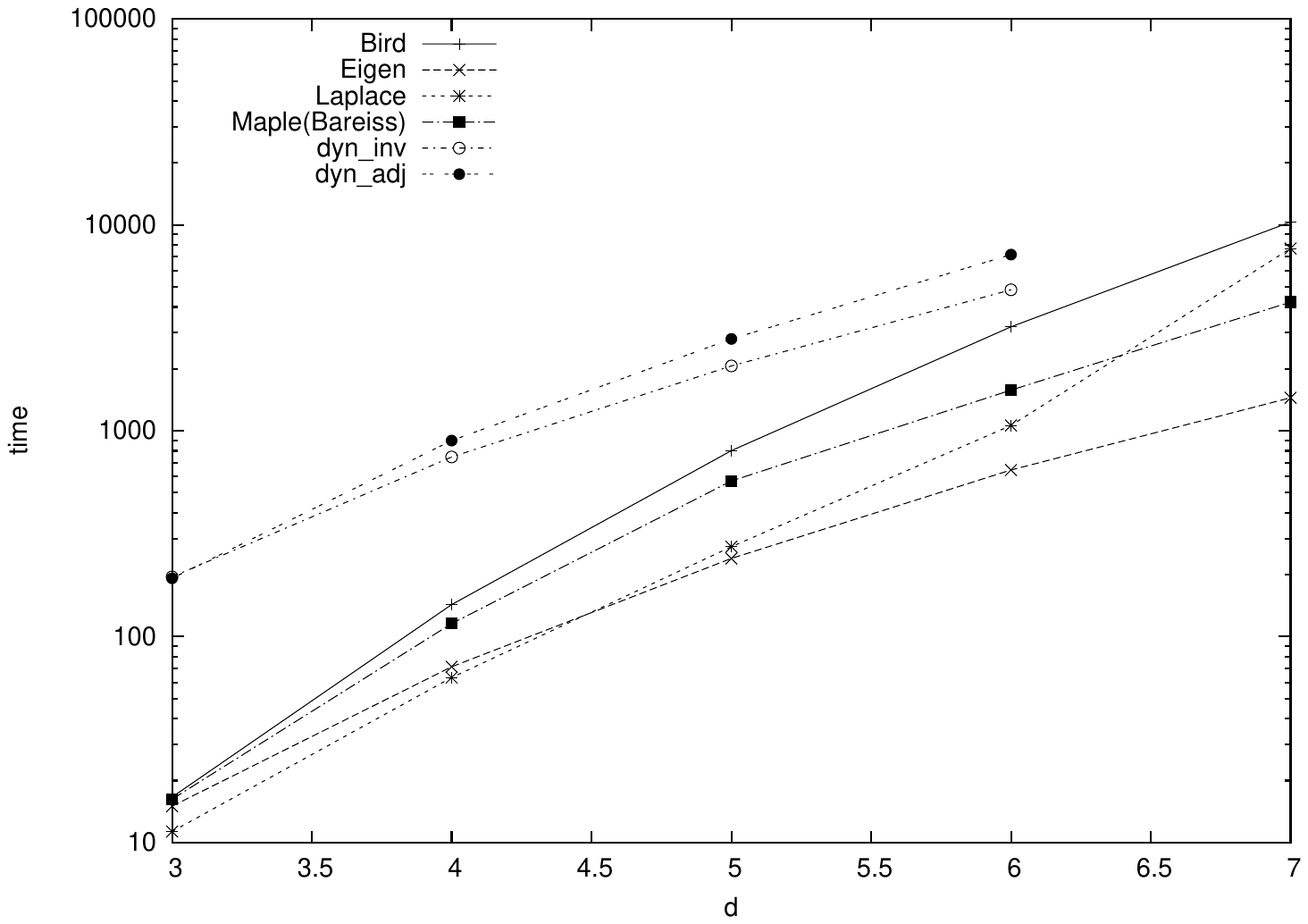}
\caption{Determinant experiments, inputs of scenario~(a). Each timing (in milliseconds) corresponds to the average of computing 10000 determinants.\label{tbl:dyndetalgos_a}}
\end{figure*}
\begin{figure*}[t]
\centering
\includegraphics[width=\textwidth]{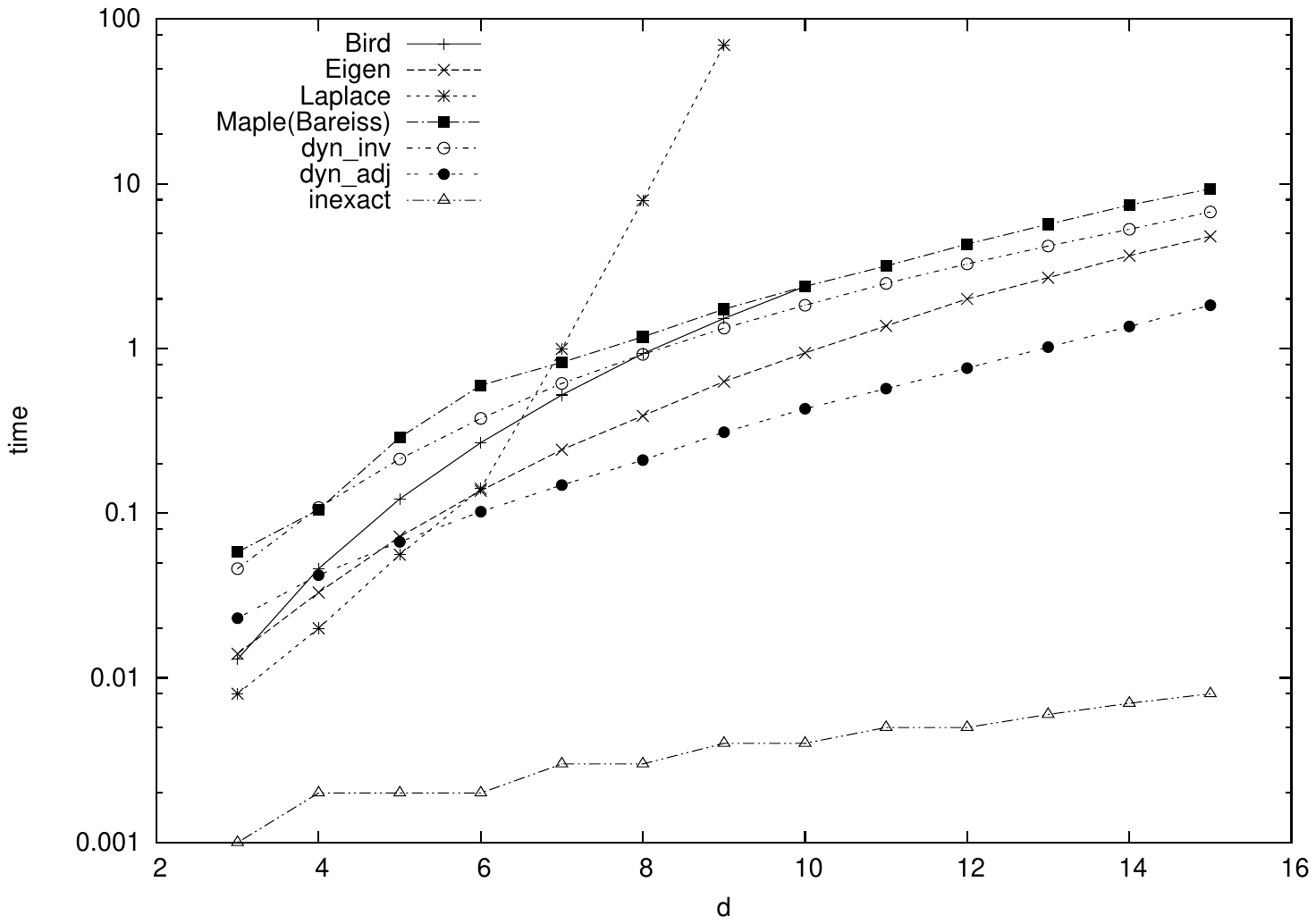}
\caption{Determinant experiments, inputs of scenario~(b). Each timing (in milliseconds) corresponds to the average of computing 10000 (for
\(d<7\)) or 1000 (for \(d \ge 7\)) determinants.\label{tbl:dyndetalgos_b}}
\end{figure*}
\begin{figure*}[t]
\centering
\includegraphics[width=\textwidth]{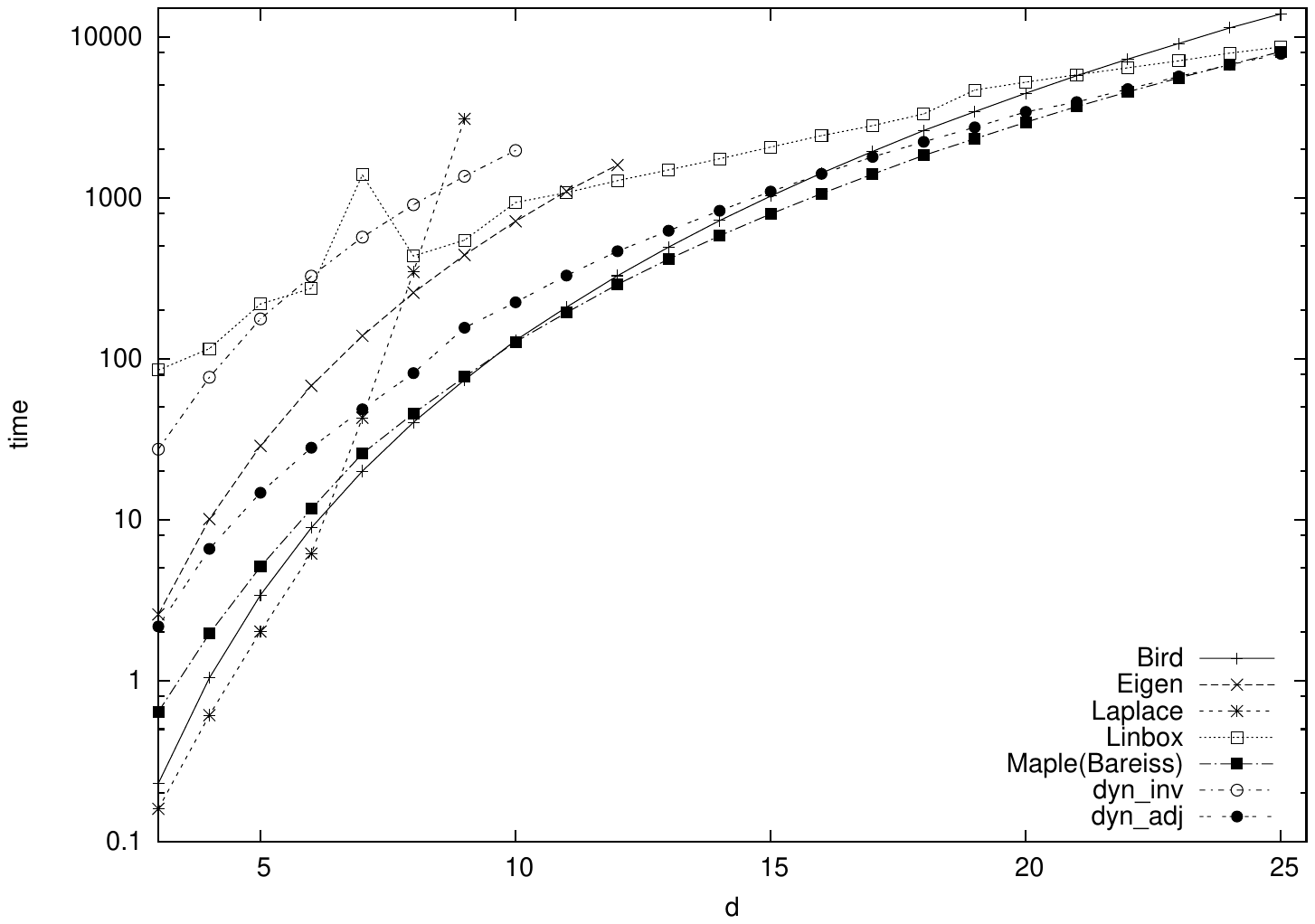}
\caption{Determinant experiments, inputs of scenario~(c).
        Times in milliseconds, averaged over
        1000 tests for \(d<9\) and 100 tests for \(d \ge 9\).\label{tbl:dyndetalgos_c}}
\end{figure*}
\begin{figure*}[t]
\centering
\includegraphics[width=\textwidth]{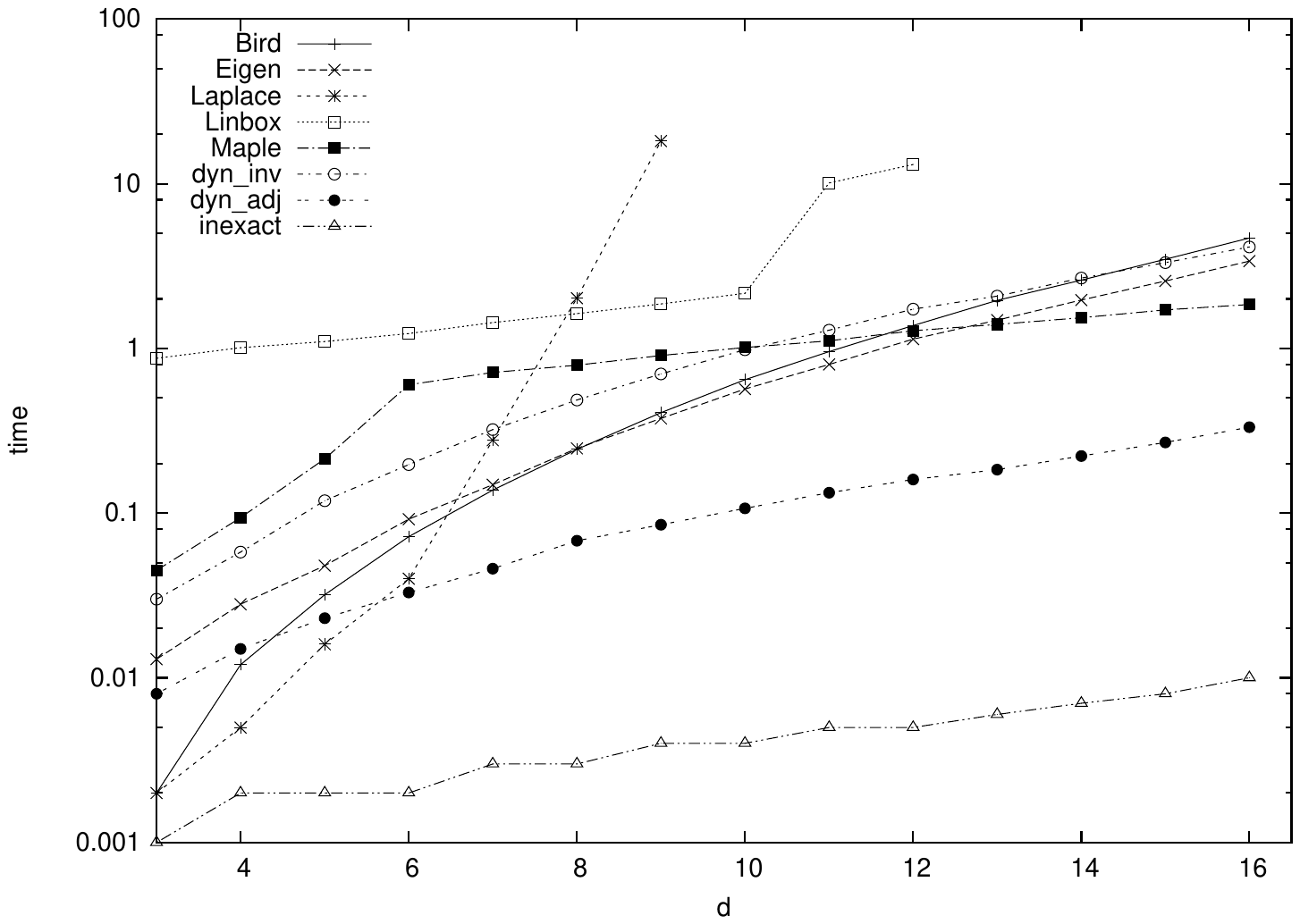}
\caption{Determinant experiments, inputs of scenario~(d). Times in
milliseconds, averaged over 10000 tests.\label{tbl:dyndetalgos_d}}
\end{figure*}

We compare state-of-the-art software for exact computation of the determinant of a $d\times d$ matrix in the four coefficient scenarios described above. When
coefficients are integers, we can use integer exact division
algorithms, which are faster than quotient-remainder division algorithms. 
In this case, division-free algorithms take advantage of using the number type {\tt mpz\_t} while the others are using {\tt mpq\_t}. 
The input matrices are constructed starting from a random $d\times d$ matrix, replacing a randomly selected column with a random $d$ vector.
We present experimental results of the four input scenarios in Figures~\ref{tbl:dyndetalgos_a}--\ref{tbl:dyndetalgos_d}.
We tested a fifth coefficient scenario (rationals of bit-size $32$), but do not show results here because timings are quite proportional to those show in Figure~\ref{tbl:dyndetalgos_a}.
We stop testing an implementation when it is slow and far from being the fastest (denoted by absence of dots in the  Figures).
On one hand, without considering the
dynamic algorithms, the experiments show the most efficient determinant algorithm
implementation in the different scenarios described.
This is a result of independent interest, and shows the efficiency of division-free algorithms in some settings.

First, we consider LU~decomposition, the current standard in 
 determinant implementations. 
We test {\tt Eigen}~\cite{eigenweb} which shown to be the fastest in scenarios (a) and (b), starting from dimension $5$ and $6$ respectively, as well as in scenario~(d) in dimensions between $9$ to $12$.

Second, we consider determinant algorithms implemented in {\tt LinBox}~\cite{DGGGHKSTV}. 
{\tt LinBox} implements state-of-the-art algorithms with the best known asymptotic complexity bounds. However, their implementation usually has a big
computational overhead and {\tt LinBox} shows the best results only when working
in high dimensions (the results of the tests of this section  corroborate this claim).
{\tt LinBox} provides a myriad of algorithms for computing determinants: many
known dense and sparse elimination methods, the block~Wiedemann
algorithm~\cite{Villard97} and an algorithm using a hybrid method mixing
Chinese remaindering and last invariant factor~\cite{DuUr05}. We tested
them and used for our tests the faster algorithm for our
scenarios~(c)~and~(d)\footnote{For technical reasons, we only tested {\tt LinBox} with integer
matrices; however, our results can be readily generalized to the rational case.}, the
hybrid elimination algorithm (which is also the default in {\tt LinBox}). 
{\tt LinBox} is never the best, due to the fact that it focuses on high
dimensions. For instance, observe
Figures~\ref{tbl:dyndetalgos_c}~and~\ref{tbl:dyndetalgos_d}. In the former,
LinBox is competitive only in high dimensions (\ie $>15$), but tends to be the most efficient in dimensions larger than 25, for which we didn't perform experiments. In the latter, {\tt LinBox} is at least two times slower than {\tt Maple} until dimension $10$. In this case, for larger dimensions, {\tt LinBox} switches the internal algorithm it uses and, while the former relation still holds, timings get much slower than {\tt Maple}.

We consider {\tt Maple}~14 {\tt LinearAlgebra[Determinant]}.
{\tt Maple} implementation chooses between Bareiss
algorithm~\cite{Bareiss68}, Gaussian elimination~\cite[\S
2.2]{DavidPoole180} and Berkowitz algorithm~\cite{Berk84}, based on the properties of the underlying algebraic structure. Note that, for scenario~(c), we experimentally check that it is  more efficient to force {\tt Maple} to use Bareiss algorithm. Experimental results of that case are presented in Figure~\ref{tbl:dyndetalgos_c}.
{\tt Maple} is the fastest only in scenario~(d), starting from dimension $13$. 

To test the behavior of the class of division-free combinatorial
algorithms, we choose to implement Bird's algorithm~\cite{Bird11} despite of the existence of  combinatorial algorithms with better asymptotic
complexity. Those algorithms are using fast matrix multiplication, which carries big constants in the complexity~\cite{I89,R05}. 
As reported in~\cite{PTVF07} implementations of fast matrix multiplications are more efficient for matrices with dimensions bigger than $100$. On the other hand, Bird's algorithm does not rely on a particular
matrix multiplication algorithm; its complexity is expressed as a function of the complexity of the matrix multiplication algorithm used.
We choose to implement Bird's algorithm using schoolbook matrix multiplication~\cite[\S 3.1]{DavidPoole180}: since Bird's algorithm only
operates with some rows of upper-triangular matrices, few scalar operations
are actually done (only \(\frac{1}{4}d^4+O(d^3)\) scalar multiplications, and the same number of additions, are needed, see~\ref{ap:bird}). 
Interestingly, naive matrix multiplication makes Bird's algorithm very competitive in small to medium dimensions. 
It is faster, in cases, than algorithms using fast matrix multiplication and faster than common decomposition methods when working with big integers.
In particular, it is the fastest in scenario~(c), starting in dimensions $7$ to $9$, and in scenario~(d), in dimensions $7$ and $8$.

The classic Laplace expansion~\cite[\S 4.2]{DavidPoole180} which falls in the category of  division-free algorithms is implemented and proved to be the most efficient until dimension 4,5,6,5 for scenario (a)-(d) respectively.
It has exponential complexity, but it behaves very well in low dimensions because of the small constant of its complexity and the fact that it performs no divisions.

We consider our implementations of \dyninv and \dynadj. In the initialization step of these algorithms, we compute the inverse, the adjoint and the determinant of the initial matrices using Gaussian elimination. This step affects only infinitesimally the total running time, because it is performed only once, and thus we did not search for optimal implementations of these algorithms.
Experiments show that \dynadj defeats the other algorithms in the most common scenarios (b), (d) starting in dimension $6$. This happens mainly because of its better asymptotic complexity.
In scenario~(c), \dynadj beats the most efficient non-dynamic methods (which are the division-free methods) only in high dimensions. It outperforms Bird only in dimension $16$, while it is faster than Bareiss only in dimension $24$.
It worth mentioning that \dynadj performs always better than
\dyninv, despite its worse arithmetic complexity. This is somehow because we are working with
multiple precision arithmetic, on which the cost of arithmetic operations is a function on the size of the operands. Since the sizes of the coefficients of the adjoint matrix are bounded, the sizes of the operands of the 
arithmetic operations in \dynadj are also bounded, which is not the case for \dyninv.

Finally, we report results of inexact computation for
scenarios~(b)~and~(d), that is, {\tt Eigen} using double-precision
floating-point arithmetic (denoted by \emph{inexact} in
Figures~\ref{tbl:dyndetalgos_b}~and~\ref{tbl:dyndetalgos_d}).  
Though largely faster than the timings of exact computations, 
the correct value of the determinant is not computed.
These experiments provide an insight of the timings one would obtain
using filtered computations, in the ideal case that no exact computation needs to be done.
See Section~\ref{subsec:fil} for a discussion on filtering.

\subsection{Geometric computation experiments}

\begin{table}
\begin{tabular*}{\textwidth}{@{\extracolsep{\fill}}c c|rr|rr|rr}
\multirow{3}{*}{$n$} & 
\multirow{3}{*}{$d$} & 
\multicolumn{2}{c|}{{\tt hdch\_q}}  & 
\multicolumn{2}{c|}{{\tt hdch\_z}}  & 
\multicolumn{2}{c}{{\tt Triangulation}}
\\
\cline{3-8}
& & \multicolumn{1}{c}{time}  & \multicolumn{1}{c|}{memory} &
    \multicolumn{1}{c}{time}  & \multicolumn{1}{c|}{memory} &
    \multicolumn{1}{c}{time}  & \multicolumn{1}{c}{memory}\\
& & \multicolumn{1}{c}{(sec)} & \multicolumn{1}{c|}{(MB)} &
    \multicolumn{1}{c}{(sec)} & \multicolumn{1}{c|}{(MB)} &
    \multicolumn{1}{c}{(sec)} & \multicolumn{1}{c}{(MB)}\\
\hline
260  &  2  &  0.02  & 35.02 &  0.01  & 33.48 &  0.05 & 35.04\\ 
500  &  2  &  0.04  & 35.07 &  0.02  & 33.53 &  0.12 & 35.08\\ 
260  &  3  &  0.07  & 35.20 &  0.04  & 33.64 &  0.20 & 35.23\\ 
500  &  3  &  0.19  & 35.54 &  0.11  & 33.96 &  0.50 & 35.54\\ 
260  &  4  &  0.39  & 35.87 &  0.21  & 34.33 &  0.82 & 35.46\\ 
500  &  4  &  0.90  & 37.07 &  0.47  & 35.48 &  1.92 & 37.17\\ 
260  &  5  &  2.22  & 39.68 &  1.08  & 38.13 &  3.74 & 39.56\\ 
500  &  5  &  5.10  & 45.21 &  2.51  & 43.51 &  8.43 & 45.34\\
\hline
260  &  6  &  14.77  & 1531.76  &  8.42    & 1132.72  &  20.01  & 55.15\\ 
500  &  6  &  37.77  & 3834.19  &  21.49   & 2826.77  &  51.13  & 83.98\\ 
220  &  7  &  56.19  & 6007.08  &  32.25   & 4494.04  &  90.06  & 102.34 \\
320  &  7  &  {\tt swap}  & {\tt swap}  &  62.01  & 8175.21 & 164.83 & 185.87\\
120  &  8  &  86.59  & 8487.80  & 45.12 & 6318.14 &  151.81 & 132.70\\ 
140  &  8  &  {\tt swap}  &  {\tt swap} &  72.81  & 8749.04 & 213.59 & 186.19\\ 
\end{tabular*}
\caption{Comparison of {\tt hdch\_q}, {\tt hdch\_z} and {\tt Triangulation}.
Points from distribution (iii) with integer coefficients; {\tt swap} means
that the machine used swap memory. Times averaged over 100 tests.\label{tbl:CHz}}
\end{table}

We perform an experimental analysis on the behavior of the application of dynamic determinants in geometric computation. 
Our main focus is to provide exact determinant constructions to volume computation. Since convex hull computation is closely connected to volume computation (cf.\ Section~\ref{sect:convex_hulls}) we study also convex hull algorithms.
We experiment with four state-of-the-art convex hull packages. 
Two of them implement incremental convex hull algorithms: {\tt Triangulation}~\cite{Hornus15} implements~\cite{CMS93} and {\tt beneath-and-beyond (bb)} implements the Beneath-and-Beyond algorithm in {\tt polymake}~\cite{Gawrilow99polymake}. 
The package {\tt cdd}~\cite{cddlib} implements the double description method, and {\tt lrs} implements the gift-wrapping algorithm using reverse search~\cite{Avis98lrs}.
All packages apart from {\tt cdd} can be used to compute volumes of  polytopes. 
We show that the application of our method  accelerates {\tt Triangulation} and outperforms other software. 

We design the input of our experiments parametrized on the number type of the coefficients and on the distribution of the points. We test our method with {\it synthetic data} first.
The number type is either rational or integer. From now on, when we refer to rational and
integer we mean scenario~(b) and (d), respectively. We test the three uniform point distributions described below. 
When the performance of the tested algorithms on two different
distributions is similar, we present the results that correspond to only
one of the distributions.
\begin{description}
\item[i.] in the $d$-cube \([-100,100]^d\), 
\item[ii.] in the origin-centered $d$-ball of radius $100$, and 
\item[iii.] on the surface of the ball of (ii).
\end{description}

\begin{figure*}[t]
\centering
\includegraphics[width=.8\textwidth]{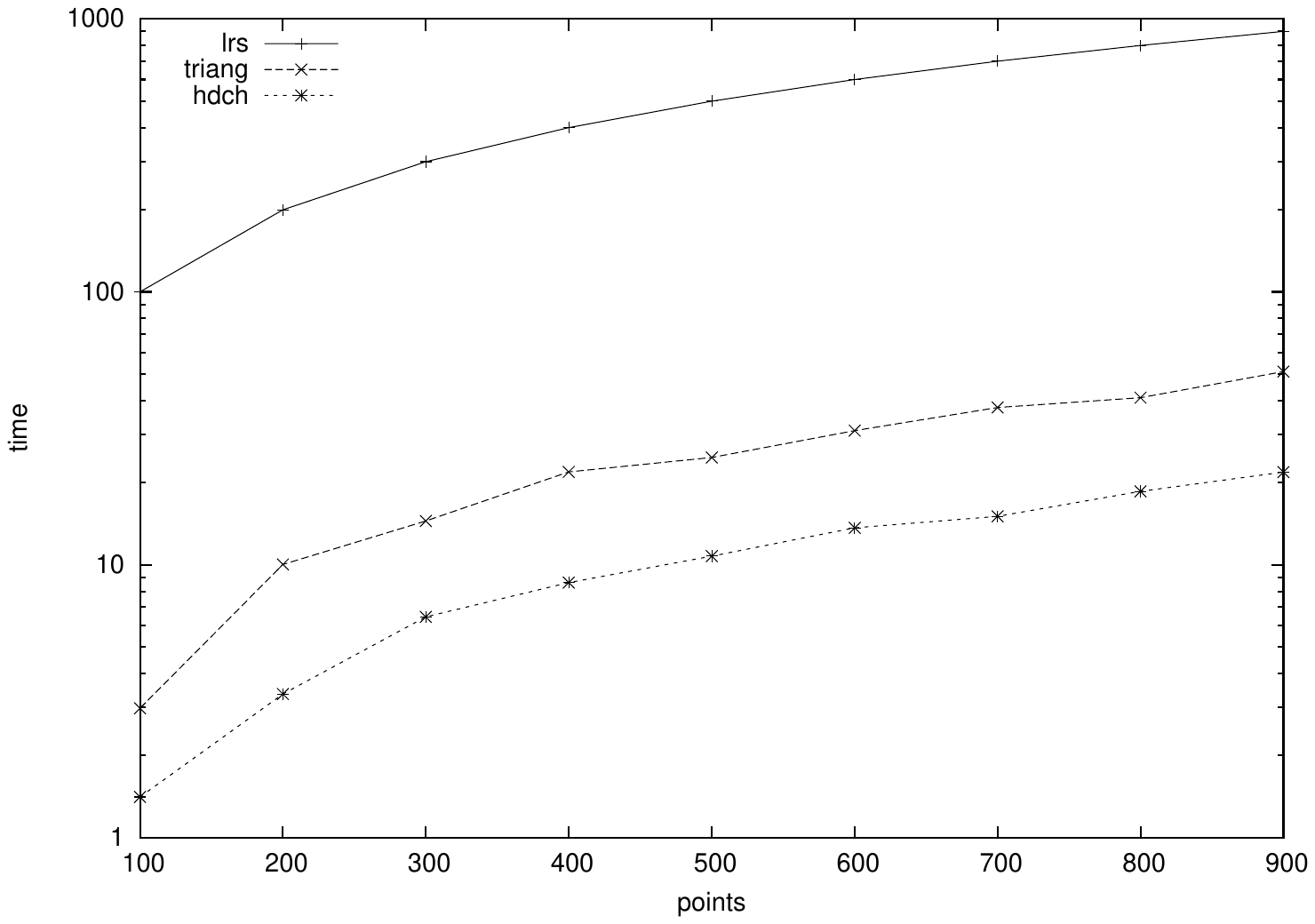}
\caption{Volume computation experiments; input is random points in a cube
of dimension $6$; \ie distribution (i). Times in seconds averaged over 100 tests.\label{tbl:vol}}
\end{figure*}

First, we test our method against volume computation provided by {\tt lrs}. Our software in dimension $6$ can be up to $20$ times faster (Figure~\ref{tbl:vol}).
This is an experimental evidence that our method could be used to compute
volumes of polytopes for which state-of-the-art methods halt.
Also note that the algorithms of {\tt vinci}~\cite{vinci} another
state-of-the-art software for exact volume computation were not faster than
{\tt lrs} in our experiments. In particular, the only available algorithm that {\tt vinci} provides
when the input polytope representation is given by points and the  inequalities are not known uses {\tt lrs}.

\begin{figure*}[t]
\centering
\subfigure[]{\label{fig:ch_ball_int_6}
\includegraphics[width=0.48\textwidth]{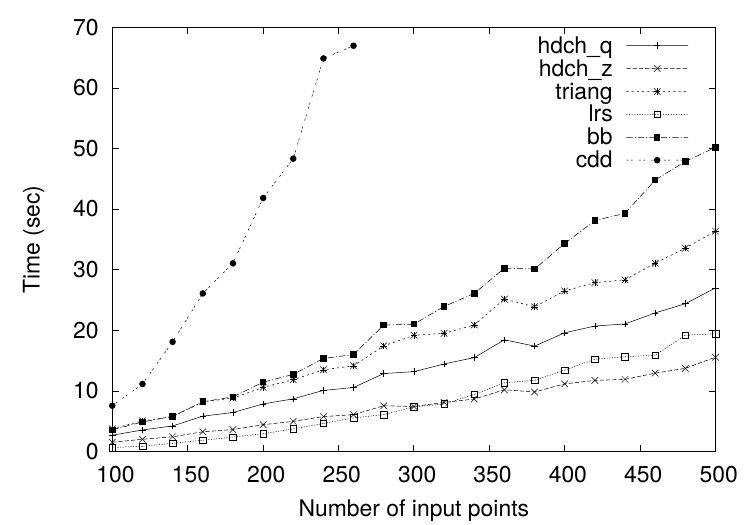}}
\subfigure[]{\label{fig:ch_sphere_int_6}
\includegraphics[width=0.48\textwidth]{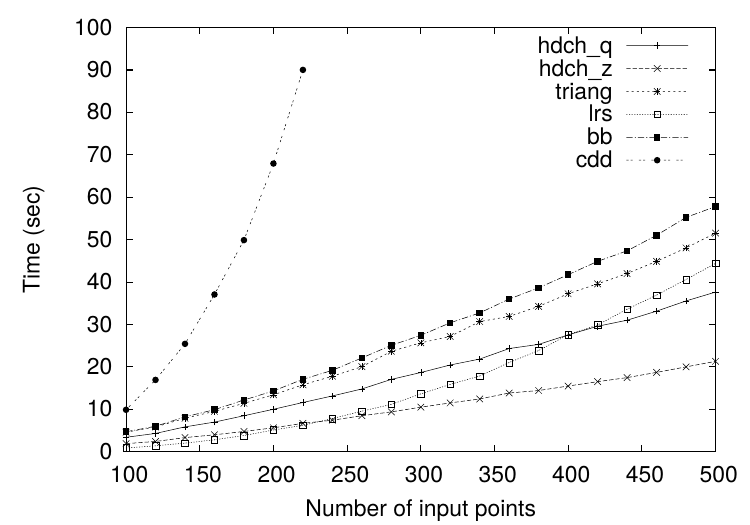}}
\caption{Comparison of convex hull packages for $6$-dimensional inputs with
integer coefficients. Points are uniformly distributed (a) inside a
$6$-ball and (b) on its surface. Times averaged over 100 tests.
\label{fig:ch}
}
\end{figure*}

\begin{figure*}[t]
\centering
\subfigure[]{\label{fig:ch_ball_rat_6}
\includegraphics[width=0.48\textwidth]{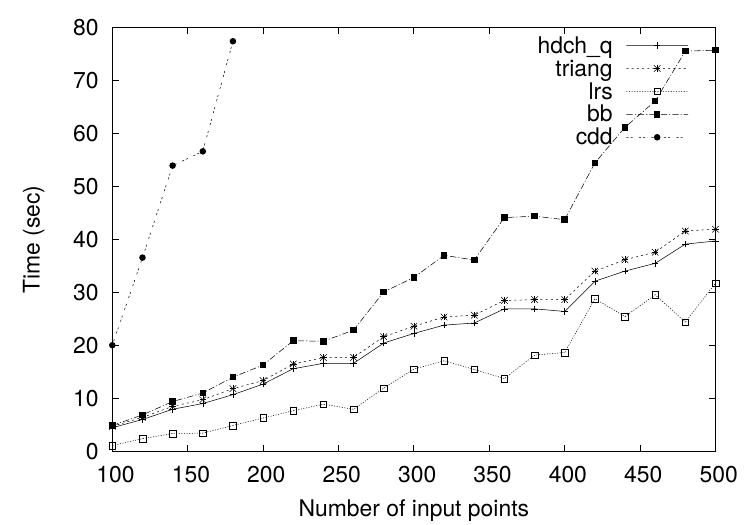}}
\subfigure[]{\label{fig:ch_sphere_rat_6}
\includegraphics[width=0.48\textwidth]{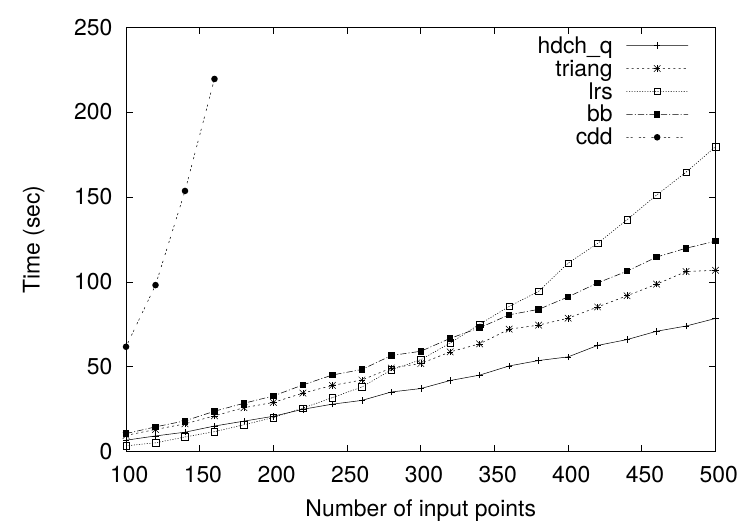}}
\caption{Comparison of convex hull packages for $6$-dimensional inputs
with rational coefficients. Points are uniformly distributed (a) inside a
$6$-ball and (b) on its surface. Times averaged over 100 tests. 
\label{fig:ch2}}
\end{figure*}

Second, we perform an experimental comparison of the four convex hull packages and {\tt
hdch}, with input points from distributions (i)--(iii) with either rational
or integer coefficients. 
In the case of integer coefficients, we test {\tt hdch} using {\tt mpq\_t} ({\tt
hdch\_q}) or {\tt mpz\_t} ({\tt hdch\_z}).
In this case {\tt hdch\_z} is the most efficient with input from
distribution (ii) (Figure~\ref{fig:ch_ball_int_6}; distribution (i) is
similar to this) while in distribution (iii) both  {\tt hdch\_z} and {\tt
hdch\_q} perform better than all the other packages (see
Figure~\ref{fig:ch_sphere_int_6}). 
In the rational coefficients case, {\tt hdch\_q} is competitive to the
fastest package (Figure~\ref{fig:ch2}).
Note that the rest of the packages cannot perform arithmetic computations
using {\tt mpz\_t} because they are lacking division-free determinant algorithms. 
It should be noted that {\tt hdch} is always faster than {\tt
Triangulation}. The sole modification of the determinant algorithm made it
faster than all other implementations in the tested scenarios. 

Moreover, we quantify the improvements of hashed dynamic
determinants scheme on {\tt Triangulation}.
For input points from distribution (iii) with integer coefficients, when
dimension ranges from $3$ to $8$, {\tt hdch\_q} is up to $1.7$ times faster
than {\tt Triangulation} and {\tt hdch\_z} up to $3.5$ times faster (see Table~\ref{tbl:CHz}). 
Table~\ref{tbl:CHz} also quantifies the memory consumption needed to obtain
these speed-ups.

We emphasize the utilization of the hashed
dynamic determinants scheme when working with {\it real data}. We carry out
experiments using as input several resultant polytopes. These polytopes are
fundamental in algebraic geometry~\cite{GKZ} and have been also studied from a  computational point of view~\cite{EFKP12journal}. The list of their applications contains polynomial system solving and computer aided design~\cite{EFKP12journal}.
A basic property of these polytopes is that their vertices have integral coefficients. 
The results in Table~\ref{tbl:res} show a speed-up of up to $3$ times
using {\tt hdch\_z} with respect to {\tt Triangulation}. The last column shows the exact volume computed for these polytopes.

\begin{table}[h]
\begin{tabular*}{\textwidth}{@{\extracolsep{\fill}}c c | r r r r}
\multirow{2}{*}{$n$} & 
\multirow{2}{*}{$d$} & 
\multicolumn{3}{c}{time (sec)} & 
\multicolumn{1}{c}{\multirow{2}{*}{volume}} \\
\cline{3-5} & &
\multicolumn{1}{c}{\tt hdch\_q} &
\multicolumn{1}{c}{\tt hdch\_z} &
\multicolumn{1}{c}{\tt Triangulation}
\\
\hline
80   &  6  &   0.54 &  0.27  &   0.66 & 368986.7\\
100  &  6  &   0.69 &  0.33  &   0.87 & 108096.3\\
110  &  6  &   1.20 &  0.52  &   1.40 & 1456226058.5\\
125  &  6  &   1.28 &  0.61  &   1.66 &  66137.3\\
376  &  7  &  17.07 &  7.80  &  24.41 & 1713149926.2\\
414  &  7  &  23.02 & 10.91  &  32.54 & 82132445.9\\
500  &  7  &  29.40 & 13.05  &  41.22 & 2593047991.6\\ 
528  &  7  &  38.22 & 17.96  &  54.91 & 33727790.7\\
\end{tabular*}
\caption{Computing resultant polytopes. Times averaged over 100 tests.\label{tbl:res}}
\end{table}

\begin{table}
\begin{tabular*}{\textwidth}{@{\extracolsep{\fill}}l c c r r r r r }
 & \multirow{3}{*}{$d$} & 
\multirow{3}{*}{$n$} &
       \multicolumn{1}{c}{preproc.}& \multicolumn{1}{c}{data}&
       \multicolumn{1}{c}{\# of}& \multicolumn{2}{c}{query time}\\
 & & & \multicolumn{1}{c}{time}& \multicolumn{1}{c}{structs.}&
 \multicolumn{1}{c}{cells in} & \multicolumn{2}{c}{(sec)}\\
       \cline{7-8}
 & & & \multicolumn{1}{c}{(sec)}& \multicolumn{1}{c}{(MB)}&
 \multicolumn{1}{c}{triangul.}& \multicolumn{1}{c}{1K}  &
 \multicolumn{1}{c}{1000K}\\
\hline
{\tt hdch\_z} & 8 & 120 & 45.20   & 6913 & 319438 & 0.41 & 392.55\\
{\tt triang} & 8 & 120 & 156.55 & 134 & 319438  & 14.42 & 14012.60\\\hline
{\tt hdch\_z} & 9 & 70 & 45.69   & 6826 & 265874 & 0.28  & 276.90\\
{\tt triang} & 9 & 70 & 176.62 & 143 & 265874 & 13.80 & 13520.43\\\hline
{\tt hdch\_z} & 10 & 50 &  43.45  & 6355 & 207190 & 0.27  & 217.45
\\
{\tt triang} & 10 & 50 &  188.68 & 127 & 207190 & 14.40  & 14453.46 \\\hline
{\tt hdch\_z} & 11 & 39 &  38.82  & 5964 & 148846 & 0.18 &   189.56 \\
{\tt triang} & 11 & 39 &  181.35 & 122 & 148846 &  14.41 & 14828.67 \\
\end{tabular*}
\caption{\label{tbl:loc}Point location experiments. Time of 1K and 1000K (1K=1000) query
        points for {\tt hdch\_z} and {\tt Triangulation} ({\tt triang}),
        using distribution (iii) for preprocessing and distribution (i) for
        queries and integer coefficients. Times averaged over 100 tests for 1K data.}
\end{table}

We test the efficiency of hashed dynamic determinants scheme on
the {\it point location} problem in a triangulation.
Given a pointset, {\tt Triangulation} constructs a triangulation of the convex hull of the pointset and a data structure that can
perform point locations of new points. 
In addition to that, {\tt hdch} constructs the hash table with matrices and determinants used for faster orientation computations.
We perform tests with {\tt Triangulation} and {\tt hdch} using input points uniformly distributed on the surface of a ball (distribution (iii)) as a preprocessing to build the data structures. Then, we perform point locations
using points uniformly distributed inside a cube (distribution (i)). 
Experiments show that our method yields a speed-up in query time by a factor of $35$ to $78$ when dimension ranges from $8$ to $11$ using points with integer coefficients (scenario~(d)) (see Table~\ref{tbl:loc}).

\subsection{Memory consumption}
The main disadvantage of {\tt hdch} is the amount of memory consumed, which
allows us to compute up to dimension $8$ (see Table~\ref{tbl:CHz}). One can
think at this point that an intelligent memory allocation scheme could
improve the performance of our algorithms. However, tests with an
implementation of {\tt hdch} using the Boehm--DeMers--Weiser conservative
garbage collector~\cite{Boehm93} did not show improvements in computing
time. This can be due to the fact that the complexity of the operations
performed on the allocated numbers surpasses the complexity of the
allocated space. Thus, changing the allocation scheme would not reduce
significantly the computation time. This drawback can be seen as the price
to pay for the obtained speed-up.

The large memory consumption of our method can be overhauled by
exploiting hybrid techniques. That is, to use the dynamic determinant hashing
scheme as long as there is enough memory and subsequently use the best
available determinant algorithm (Section~\ref{sect:experiments}). 
Alternative options are to clean periodically the hash table or to use a 
Least Recently Used (LRU) cache
to avoid storing for long time unused determinants and matrices. 
For the latter, techniques for efficiently computing determinants of
matrices with more than one update, as described by
Sankowski~\cite{Sankowski04}, could be utilized.

\subsection{Filtering}
\label{subsec:fil}

As shown by experiments, one main advantage of the dynamic determinant
method shows up when applied to exact geometric constructions. One question
that arises, and could be a subject of future work, is whether we can use
this method to geometric predicates that benefit from filtering techniques.  
While in low dimensions filtering provides a very efficient framework for
computing signs of determinants, in medium and high dimensions filtering
with double-precision floating-point numbers is not efficient, since it
reverts too often to exact computations~\cite{BoiDevHor09}.
Recent work in {\tt CGAL}, namely the {\tt Epick\_d} kernel, tries to overcome this limitation using hardware and software advances, pushing forward the dimensions on which filtering can be used. 
Preliminary tests indicate that our implementation, without filtering, is
faster than the filtering implemented by
Boissonnat~\etal~\cite{BoiDevHor09}, but slower than the new implementation
in {\tt Epick\_d}.

Br{\"o}nnimann~\etal~\cite{BroBurPio01} propose another filtering scheme for determinant computations in medium dimensions, using a decomposition method which is numerically more stable than the usual LU~decomposition. However, the authors are not aware of any work that evaluates the efficiency of this technique in practice. 

\section{Concluding remarks}

Our paper introduces a method of optimizing the computation of sequences of determinants, using dynamic determinant updates and the well-known Sherman--Morrison formulas. Despite of being well-known this is the first time these formulas are use to geometric algorithms, which make heavy use of similar determinant computations. We demonstrate how this can be done and also present experimental evidences about the supremacy of these methods over state-of-the-art methods in determinant and geometric computations.  

A future improvement in the memory consumption of our method could be 
the exploitation of hybrid memory management techniques as discussed in 
Section~\ref{sect:experiments}.
One extension of the proposed method of this paper would be the application of dynamic determinants to the {\it gift wrapping} (GfR) convex hull algorithms~\cite{CK70,Avis98lrs}. 
Such an extension would certainly improve the memory consumption of our method. 

Overall we hope that the research results presented in this paper will promote the use of these update formulas in other geometric algorithms implementations, and will trigger some further applied-research regarding searching and storing the determinant-adjoint pairs as well as the use of dynamic determinants methods together with filtered computations.

\section*{Acknowledgements}
We would like to thank Ioannis~Z.~Emiris for his advice and encouragement,
Elias~Tsigaridas for bibliographic suggestions, Menelaos~Karavelas for
discussions on efficient dynamic determinant updates, Olivier~Devillers for
discussions on {\tt Triangulation} and Monique~Teillaud for her careful
comments on a previous version of the manuscript.
We also thank the anonymous referees for their comments which helped us improve the presentation.
This work is partially supported by project ``Computational Geometric
Learning'', which acknowledges the financial support of the Future and
Emerging Technologies (FET) programme within the Seventh Framework
Programme for Research of the European Commission, under FET-Open grant
number: 255827. The main part of this work was done while the authors were
at the University of Athens.

\bibliographystyle{abbrv} 
\bibliography{bibliography}

\pagebreak

\appendix

\section{Complexity of Bird's determinant algorithm}
\label{ap:bird}

So far we choose to implement Bird's algorithm~\cite{Bird11} to
represent the class of combinatorial determinant algorithms. In the
original paper, it is stated that the complexity is bounded by
\(O\big(dM(d)\big)\), where \(M(d)\) is the cost of multiplicating two
matrices.

We choose in this paper to implement the above mentioned algorithm using
schoolbook matrix multiplication~\cite[\S 3.1]{DavidPoole180}. The given
complexity bound still holds, but we compute in this section a tighter
bound for our specific case.

The tool we use in the analysis is Faulhaber's formula~\cite{Con96}, which
gives a form to compute sums of powers.
For particular values of the exponent of the summed numbers,
this formula turns into
\begin{align}
        \label{eq:sum2}
        P(d) &= \sum_{k=1}^d k^2 = \frac16 \big( 2d^3 + 3d^2 + d \big)
                \text{, and}\\
        \label{eq:sum3}
        T(d) &= \sum_{k=1}^d k^3 = \frac14 \big( d^4 + 2d^3 + d^2 \big)
                \text{.}
\end{align}

With Equations~\ref{eq:sum2}~and~\ref{eq:sum3}, we are ready to develop the
formulas to compute the complexity bound. Let us assume that the matrix has
size $d$. We will compute the number of scalar multiplications done by the
algorithm, and show then that the number of additions is bounded by the
same function.

The algorithm performs, on its first step, a partial multiplication of one
upper-triangular matrix and a full matrix. Moreover, only the upper
triangular part of this matrix will be used in the next step; thus we
consider in the sequel only the computation of the entries which will be
used.

Let us consider the rows of the resulting matrix. The first row contains
$d$ elements and, to compute each one of them, we need $d$ multiplications.
The second row contains $d-1$ non-zero elements and, to compute each one of
them, we need to perform only $d-1$ multiplications (since we do not
perform one multiplication, because we know the first element of the second
row of an upper-triangular matrix is zero). With analogous reasoning for
each row, we can conclude that, to compute the $(d-k)$-th row of the
product, we need $k^2$ multiplications. To compute the first matrix, we need
thus $P(d)=\sum_{k=1}^d k^2$.

Let us consider the second step. Of the matrix computed on the first step,
we need all but the last row. In fact, in step $s$ of the algorithm, we
need only the first $d-s$ rows. In fact, we will compute only the rows
which are needed. This means that, in step $s$ of the algorithm, we will
perform $P(d)-P(s)$ scalar multiplications. It follows that the amount of
scalar multiplications needed by the algorithm is
\begin{equation}
        \label{eq:tn}
        A(d) = d P(d) - \sum_{j=0}^{d-1} P(j)
\end{equation}

Since we know how to compute the minuend, let us concentrate on the sum
of the $P(j)$'s.

\begin{align}
        \nonumber
        \sum_{j=0}^{d-1} P(j) &=
                \frac13 \Bigg(\sum_{j=0}^{d-1} j^3 \Bigg) +
                \frac12 \Bigg(\sum_{j=0}^{d-1} j^2 \Bigg) +
                \frac16 \Bigg(\sum_{j=0}^{d-1} j \Bigg)\\
        \nonumber
        &=
                \frac13 T(d-1) +
                \frac12 P(d-1) +
                \frac1{12} (d-1)d\\
        \nonumber
        &=
                \frac1{12} \Big(
                        (d-1)^4 + 2(d-1)^3 + (d-1)^2 +
                        2(d-1)^3 + 3(d-1)^2 + (d-1) +
                        (d-1)d
                \Big)\\
        \label{eq:sumcj}
        &=
                \frac1{12} \Big(
                        (d-1)^4 + 4(d-1)^3 + 4(d-1)^2 + (d-1)(d+1)
                \Big)
\end{align}

Substituting Equation~\ref{eq:sumcj} in Equation~\ref{eq:tn} we have the
following.

\begin{align}
        \nonumber
        A(d) &= d P(d) - \sum_{j=0}^{d-1} P(j)\\
        \nonumber
        &= \frac1{12} \Big( 4d^4 + O(d^3) \Big) -
        \label{eq:multiplications}
        \frac1{12} \Big( (d-1)^4 + O(d^3) \Big)\\
        &= \frac14 d^4 + O(d^3)
\end{align}

Equation~\ref{eq:multiplications} bounds the number of scalar
multiplications done by Bird's algorithm when using schoolbook matrix
multiplication. Let us now bound the number of scalar additions done.
Observe that, for multiplicating two matrices, the number of scalar
additions is always smaller than the number of scalar multiplications.
Beyond those, the algorithm needs to perform at most $d$ scalar additions
on each step. This means that the number of scalar additions performed by
the algorithm is also bounded by $A(d)$.

\end{document}